\documentclass[11pt,a4paper]{article}
\usepackage{fullpage}
\usepackage{amsmath, amsthm, amsfonts,  amssymb, latexsym}
\usepackage[colorlinks, citecolor=Red, linkcolor=Blue, urlcolor=blue]{hyperref}
\usepackage{caption}
\usepackage[dvipsnames]{xcolor}
\usepackage{tikz}
\usepackage{mcite}
\usepackage{enumerate}
\usepackage{pgfplots}

\usetikzlibrary{patterns}
\newcommand{\rhs}{r.h.s.\ }

\newcommand{\wrt}{w.r.t.\ }
\newcommand{\cf}{cf.\ }

\newcommand{\bra}[1]{\langle #1 \rvert}
\newcommand{\ket}[1]{\lvert #1 \rangle}
\newcommand{\braket}[2]{\langle #1 \vert #2 \rangle}

\newcommand{\ud}{\mathrm{d}}
\newcommand{\del}{\partial}

\newcommand{\R}{\mathbb{R}}

\newcommand{\order}{\mathcal{O}}

\newcommand{\cP}{\mathcal{P}}

\newcommand{\cH}{\mathcal{H}}

\newcommand{\eps}{\varepsilon}

\newcommand{\ri}{{\mathrm{in}}}
\newcommand{\WKB}{{\mathrm{WKB}}}

\renewcommand{\bf}[1]{{\mathbf{#1}}}

\newcommand{\nn}{\nonumber}
\newcommand{\beq}{\begin{equation}}
\newcommand{\eeq}{\end{equation}}

\newcommand{\defeq}{\mathrel{:=}}
\newcommand{\eqdef}{\mathrel{=:}}

\newcommand{\fin}{{\mathrm{fin}}}
\newcommand{\delay}{{\mathrm{delay}}}
\newcommand{\class}{{\mathrm{class}}}

\DeclareMathOperator{\Real}{Re}
\DeclareMathOperator{\Imag}{Im}

\newcommand{\cO}{{\mathcal{O}}}

\usepackage[T1]{fontenc}
\begin{document}

\title{Angular time delay in quantum mechanical scattering}

\author{Jochen Zahn \\ Institut f\"ur Theoretische Physik, Universit\"at Leipzig \\ Br\"uderstr.\ 16, 04103 Leipzig, Germany \\ jochen.zahn@itp.uni-leipzig.de}

\date{\today}

\maketitle

\begin{abstract}
We apply Brunetti and Fredenhagen's concept of the time of occurrence of an event in quantum mechanics \cite{BrunettiFredenhagenTimeObservable} to the example of scattering off a spherical potential. Thereby, we re-derive the expression of Froissart, Goldberger, and Watson for the angular time delay \cite{FroissartGoldbergerWatson}, clarifying some conceptual issues with their derivation. We also present an elementary re-derivation of the ``space shift'' (essentially the impact parameter) defined in the quantum mechanical context by the same authors. We clarify the relation of both quantities to their classical counterparts in the context of the WKB approximation. As an example, we apply the concepts to scattering at a hard sphere. We find pronounced peaks in the both the time delay and the space shift at the minima of intensity in the forward diffraction region for short wavelength scattering and discuss whether these could in principle be observable.
\end{abstract}

\section{Introduction}

The problem of time observables in quantum mechanics is a topic nearly as old as quantum mechanics itself, \cf the Introduction of \cite{TimeInQuantumMechanicsI} for a historical overview. In the traditional formulation of quantum mechanics, time is not an observable, but a parameter. In other words, the formalism answers questions like: ``When measuring observable $A$ at time $t$ on a system prepared in the state $\Psi$ at time $t_0$, what is the probability of measuring the value $a$?'' However, not all experimental situations are of this type. This is the case in particular for scattering experiments, where the outgoing particles are registered by detectors which are constantly turned on. In such a situation, the question ``When does the detector click?''~\cite{BrunettiFredenhagenTimeObservable} seems to be meaningful and experimentally accessible, at least in principle. It is thus probably no coincidence that the first definition of a quantity closely related to a time observable was given in the context of scattering, namely the Eisenbud-Wigner time delay \cite{Eisenbud}
\beq
\label{eq:Eisenbud}
 t_{\delay, \ell} = 2 \del_E \delta_\ell(E)
\eeq
with $\delta_\ell$ the scattering phase of the $\ell$th partial wave.\footnote{Throughout, we are using units in which $\hbar = 1$.} The relation \eqref{eq:Eisenbud} is derived in \cite{Wigner55} using stationary phase arguments. 

It is important to note that the physical interpretation of the time delay of a partial wave is a priori unclear, as an incoming wave consists of a superposition of partial waves and a typical detector discriminates between different scattering angles, not angular momenta (it can however be justified in a high energy approximation, see Section~\ref{sec:WKB} below). However, the relation \eqref{eq:Eisenbud} is straightforwardly adjustable to the case of one-dimensional scattering at a potential barrier, where $2 \delta_\ell$ should be replaced by the phase of the transmission amplitude. In that case, the interpretation is clear: It is the time delay \wrt the situation in the absence of the potential. Furthermore, as the derivation is based on stationary phase arguments, it should be interpreted as an expectation value.

For the case of three-dimensional scattering, a time delay for the full scattering process, not just partial waves, was given by Goldberger, Froissart, and Watson \cite{FroissartGoldbergerWatson}:\footnote{See also \cite{Smith1960, GoldbergerWatson}, which contained the definition somewhat implicitly, and without derivation.}
\beq
\label{eq:GoldbergerFroissartWatson}
 t_\delay(\theta) = \del_E \arg A(E, \theta).
\eeq
Here $A(E, \theta)$ is the scattering amplitude at energy $E$ and angle $\theta$. This should be interpreted as the time delay as measured by the detector at scattering angle $\theta$. Again, this expression is derived using stationary phase arguments.

Driven by the availability of attosecond light pulses, the topic of time delay in photoemission has attracted considerable attention in recent years, \cf \cite{SchultzeEtAl, DahlstromIntroduction} for example. Also its angular dependence has been studied extensively in recent years, both theoretically and experimentally, see \cite{WaetzelEtAl, heuser2016angular, ivanov2017angle, Amusia, Aiswarya}, for example.
The angular time delay \eqref{eq:GoldbergerFroissartWatson} was also suggested as a tool to identify resonances in reactive molecular collisions \cite{ChaoSkodje}.
Recently, the partial wave time delay \eqref{eq:Eisenbud} was used to discuss causality (violation) in gravitational effective theories \cite{deRhamCausalityConstraints}.

The interpretation of \eqref{eq:GoldbergerFroissartWatson} as a time delay has been critized \cite{Nussenzveig72} on two grounds: First, the derivation based on the stationary phase arguments does not take into account the reshaping of the wave packet due to scattering. Second, the expression can not be interpreted as a time delay, because in the absence of a scattering potential, no scattering to angle $\theta$ occurs, so it is unclear \wrt which reference process it is defined. The main objective of this paper is to argue that these objections can be refuted.\footnote{We note that in the alternative treatment proposed in \cite{Nussenzveig72}, which is based on the dwell time of the particle inside a ball around the scatterer, the time delay \eqref{eq:GoldbergerFroissartWatson} also occurs, in an integral over the angles, weighted with the differential cross section. This also suggests the interpretation of \eqref{eq:GoldbergerFroissartWatson} as the time delay at scattering angle $\theta$, but does not prove it.} We use a formalism proposed by Brunetti and Fredenhagen \cite{BrunettiFredenhagenTimeObservable} for the definition of positive operator valued measures describing the distribution of occurrence times of general ``effects'', described by projectors (or, more generally, positive operators). This formalism provides an idealized description of the measurement, without the need to introduce a dissipative mechanism. The determination of the full distribution of occurrence times requires the knowledge of the wave function $\psi(E)$, which can in general not be assumed. However, if either a reference effect or a reference dynamics is available, then the difference of the first moments of the distribution, i.e., the time delay, only depends on $| \psi(E) |^2$, i.e., the relative phases between the $\psi(E)$ at different energies $E$ become irrelevant. In particular, for $\psi(E)$ peaked at some energy $E$, it suffices to compute the time delay at this energy. No stationary phase arguments are needed.

Regarding the second objection (absence of a reference process), two approaches are possible. One could take the detector at a reference angle $\theta_0$ as a reference effect, in which case one obtains
\beq
\label{eq:t_delay_reference_angle}
 t^{\theta_0}_\delay(\theta) = t_\delay(\theta) - t_\delay(\theta_0),
\eeq
where the expressions on the \rhs are defined by \eqref{eq:GoldbergerFroissartWatson}. The second option is to use a reference dynamics, in which case a natural choice is scattering at a point scatterer, i.e., a hard sphere of radius $R$ in the limit $R \to 0$.\footnote{This was, somewhat implicitly, also the reference process used in \cite{FroissartGoldbergerWatson}.} With this reference dynamics, one obtains the time delay~\eqref{eq:GoldbergerFroissartWatson}, as shown below.

In \cite{FroissartGoldbergerWatson}, also an expression for the ``space shift'', i.e., the displacement of the outgoing particle from the axis of symmetry (essentially the impact parameter), is given. In fact, this expression is derived alongside \eqref{eq:GoldbergerFroissartWatson}, i.e., using wave packets and stationary phase arguments. We will present a considerably simpler re-derivation of this result, which evades the use of wave packets and also indicates how this quantity can in principle be measured.

As both the angular time delay and the space shift have counterparts in classical scattering, it is important to clarify the relation of the quantum mechanical concepts to the classical ones. We show that in the WKB approximation, both quantities assume their classical values, at least when classically only a single impact parameter contributes to scattering at a given angle $\theta$.

When evaluating angular time delay and space shift for scattering at a hard sphere, we find, in the long wavelength regime, essentially angle independent time delay and phase shift, as expected for s wave dominated scattering. At short wavelengths, the quantum results approach the classical ones, but there are pronounced oscillations both of time delay and phase shift in the forward diffraction region. 

The article is structured as follows: In the next section, we describe in detail the formalism of Brunetti and Fredenhagen \cite{BrunettiFredenhagenTimeObservable}. In particular, we show that in the limit of an infinitesimally thin detector, one recovers Kijowski's distribution \cite{Kijowski74}. In Section~\ref{sec:1d}, we use the framework to discuss scattering in one dimension. In particular, we exemplify the two possible ways to define a time delay, with reflection time delay (via a reference effect) and transmission time delay (via a reference dynamics). The case of scattering at a spherical potential is discussed in Section~\ref{sec:3d}. In particular, we recover the expression \eqref{eq:GoldbergerFroissartWatson} for the time delay of scattering at scattering angle $\theta$. In Section~\ref{sec:SpaceShift} we present an elementary derivation (and provide an interpretation) of the ``space shift''. In Section~\ref{sec:WKB}, we show that in the WKB approximation, both the time delay and the space shift reduce to their classical counterparts, provided that classically only a single impact parameter contributes to scattering at angle $\theta$. In Section~\ref{sec:Examples}, we apply the formalism to scattering at a hard sphere and qualitatively discuss the results. We conclude with a summary and an outlook.

\section{Time of occurrence of an event}
\label{sec:BrunettiFredenhagen}

In quantum mechanics, one describes a detector, i.e., a measuring device answering yes-no questions, by a projector $P$. The prototypical example is a projector $P_V \ket{\bf x} = \chi_V(\bf x) \ket{\bf x}$ on a region $V$ in space, describing a detector sensitive to a particle in that region (more general detectors also sensitive to the direction of the momentum of the particle will be discussed in the concrete examples in the next two sections). One may then ask: ``At what time does the detector click?''. A formalism to answer this question without the need to model the specifics of the detector
(for example by some dissipative mechanism, \cf \cite{MugaLeavensReview, RuschhauptMugaHegerfeldt} for reviews)  
has been proposed by Brunetti and Fredenhagen \cite{BrunettiFredenhagenTimeObservable}. Let $P_t = e^{i H t} P e^{- i H t}$ be the time-evolution of $P$ with the Hamilton operator $H$ (we are working in the Heisenberg picture). One interprets
\beq
 P(\R) \defeq \int_{\R} P_t \ud t
\eeq
as the observable for the total time the particle spends in the detector. Its expectation value coincides with the notion of \emph{dwell time}, \cf \cite{nussenzveig1969causality} and the review \cite{CarvalhoNussenzveigReview}, with the difference that we are here concerned with the dwell time inside the detector, not inside some region containing the scatterer. The Hilbert space now decomposes in the direct sum
\beq
 \cH = \cH_{\fin} \oplus \cH_0 \oplus \cH_\infty,
\eeq
where states in $\cH_\fin$ have a finite expectation value for $P(\R)$, whereas it vanishes (diverges) for states in $\cH_0$ ($\cH_\infty$). Elements of $\cH_0$ are states for which the effect never takes place. Examples for states in $\cH_\infty$ are bound states. A probability distribution for the time of the occurrence of the event described by $P$ can be defined for states in $\cH_\fin$. Namely, one defines the positive-operator valued measure (POVM)
\beq
\label{eq:cP_I}
 \cP^P(I) \defeq P(\R)^{- \frac{1}{2}} \left[ \int_I P_t \ud t \right] P(\R)^{-\frac{1}{2}}
\eeq
on $\cH_\fin$ for the distribution of arrival times. Here $I$ is a time interval (or a union thereof).
The first moment of this measure, i.e.,
\beq
\label{eq:TimeObservable}
 T^P \defeq \int_\R t \cP^P(\ud t),
\eeq
is then an operator such that $\bra{\Psi} T^P \ket{\Psi}$ is the expectation value (in the statistical sense) for the time of occurrence of the event $P$ in the state $\ket{\Psi} \in \cH_\fin$.
One may also relax the requirement of $P$ being a projector and allow for $P$ being a positive operator.

We emphasize that for any $\Psi \in \cH_\fin$, \eqref{eq:cP_I} provides the full distribution of click times, not just an expectation value. As we will see below, in the case of an infinitesimally thin detector and when restricting to positive momenta in the direction normal to the detector, one recovers Kijowski's distribution \cite{Kijowski74}.  In many cases of practical interest, the state $\Psi$ is not fully known, but only partial information is available, such as being peaked at some energy $E$. Then, the formalism may still be predictive if we restrict to the expectation value $T^P$, or rather differences of such expectation values, i.e., \emph{time delays}.

In the following, we will work on a subspace $\cH_{\mathrm{scat}} \subset \cH_\fin$ of scattering states $\ket{\Psi}$ which are fully characterized by a wave function $\psi(E)$ depending only on the energy, i.e.,
\beq
\label{eq:psi_E}
 \ket{\Psi} = \int \psi(E) \ket{E} \ud E,
\eeq
with $\{ \ket{E} \}$ providing a basis of eigenstates of energy $E$. For the matrix elements of $P_t$ in this basis, we have
\beq
 \bra{E'} P_t \ket{E} = e^{- i (E - E') t} P(E', E),
\eeq
with $P(E', E)$ the matrix elements of $P$. It follows that the matrix elements of $P(\R)$ are given by
\beq
 P(\R)(E', E) = 2 \pi \delta(E- E') P(E', E),
\eeq
so that
\beq
 \cP^P(I)(E', E) = \frac{1}{2 \pi} \int_I e^{- i (E-E') t}c^P(E', E) \ud t,
\eeq
where we have introduced the integral kernel
\beq
 c^P(E', E) \defeq P(E', E')^{-\frac{1}{2}} P(E', E) P(E, E)^{- \frac{1}{2}},
\eeq 
which fulfills the normalization condition
\beq
 c^P(E, E) = 1.
\eeq

Let us apply this in a concrete example to verify that for a particular choice of the effect $P$, we recover Kijowski's distribution \cite{Kijowski74}. We restrict to the one-dimensional case and, as Kijowski, positive momenta. The generalized energy eigenstates $\ket{E}$ are then given by
\beq
\label{eq:x_E_free}
 \braket{x}{E} = \sqrt{ \frac{m}{2 \pi k} } e^{i k x}
\eeq
in position space, with
\beq
\label{eq:def_k}
 k = \sqrt{2 m E}.
\eeq
We place a detector of width $\delta$ at the origin, i.e.,
\beq
 P = \chi_{[0, \delta]}(x).
\eeq
It follows that
\beq
 P(E', E) = \frac{m}{2 \pi \sqrt{k k'}} \int_0^\delta e^{i (k - k') x} \ud x = \frac{m \delta}{2 \pi \sqrt{k k'}} + \cO(\delta^2).
\eeq
In the limit $\delta \to 0$, we thus obtain
\beq
 c^P(E', E) = 1,
\eeq
which entails
\beq
 \bra{\Psi} \cP^P([t, t + \ud t]) \ket{\Psi} = \left[ \frac{1}{2 \pi} \int \ud E \ud E' \ \bar \psi(E') \psi(E) e^{- i (E - E') t} \right] \ud t.
\eeq
The expression in square brackets on the \rhs can easily be seen to coincide with Kijowski's probability density \cite{Kijowski74,RuschhauptMugaHegerfeldt}
\beq
 \Pi(t) = \frac{1}{m} \bra{\Psi(t)} \hat k^{\frac{1}{2}} \delta( \hat x ) \hat k^{\frac{1}{2}} \ket{\Psi(t)}.
\eeq

To obtain the expectation value of the time observable \eqref{eq:TimeObservable}, we restrict to wave functions $\psi(E)$ vanishing rapidly enough for $E \to 0$ and $E \to \infty$, and compute
\begin{align}
 \bra{\Psi} T^P \ket{\Psi} & = \frac{1}{2 \pi} \int t e^{- i (E-E') t} \bar \psi(E') c^P(E', E) \psi(E) \ud t \ud E \ud E' \nn \\
 & = \frac{1}{2 \pi} \int e^{- i (E-E') t} (- i \del_E) ( \bar \psi(E') c^P(E', E) \psi(E)) \ud t \ud E \ud E' \nn \\
\label{eq:Psi_T_P_Psi}
 & = \int \bar \psi(E) ( - i \del_E + t^P(E) ) \psi(E) \ud E,
\end{align}
with 
\beq
 t^P(E) \defeq - i \del_E c^P(E', E)|_{E' = E}.
\eeq

To evaluate \eqref{eq:Psi_T_P_Psi}, we need to know, due to the presence of $\del_E$, the relative phase between the amplitudes $\Psi(E)$ at different (nearby) energies $E$. Such information is typically not available in scattering experiments. There are two possibilities to overcome this difficulty, which both rely on considering differences of two expressions of the form \eqref{eq:Psi_T_P_Psi}, such that the terms involving $\del_E$ cancel. One possibility is to consider two different events $P_1$ and $P_2$ and the difference of the corresponding expectation values \eqref{eq:Psi_T_P_Psi}. If one also assumes that $\psi(E)$ is sharply peaked at energy $E$, one finds
\beq
 \bra{\Psi} (T^{P_1} - T^{P_2}) \ket{\Psi} = t^{P_1}(E) - t^{P_2}(E) \eqdef t_\delay
\eeq
which does not depend on further detailed knowledge of the wave function $\psi(E)$. This is the time delay between two events.

The second option is to consider two different Hamiltonians, i.e., time evolutions. For this, one has to match the energy eigenstates corresponding to the different Hamiltonians. In a scattering situation this proceeds by choosing the same ``in'' part in the definition of $\ket{E}$ for the different Hamiltonians. The resulting difference between the corresponding expectation values \eqref{eq:Psi_T_P_Psi} is the time delay between two evolutions. Both notions of time delay will be exemplified in the next section.

\section{One-dimensional scattering}
\label{sec:1d}

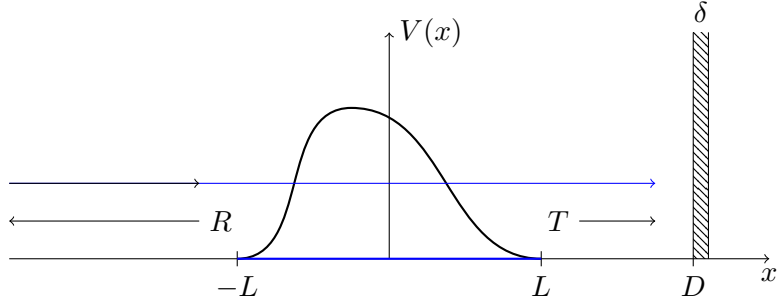
\begin{figure}
\centering
\begin{tikzpicture}
\draw[->] (-5,0) -- (5,0) node[below]{$x$};
\draw[->] (0,0) -- (0,3) node[right]{$V(x)$};
\draw (-2,0.1) -- (-2,-0.1) node[below]{$-L$};
\draw (2,0.1) -- (2,-0.1) node[below]{$L$};
\draw[thick] (-2,0) to[out=0,in=180] (-0.5,2) to[out=0,in=180] (2,0);
\draw[thick,blue] (-2,0) -- (2,0);
\fill[pattern=north west lines] (4,0) -- (4,3) -- (4.2,3) -- (4.2,0) -- (4,0);
\draw (4,-0.1) node[below]{$D$} -- (4,3);
\draw (4.2,0) -- (4.2,3);
\draw (4.1,3) node[above]{$\delta$};
\draw[blue,->] (-5,1) -- (3.5,1);
\draw[->] (-5,1) -- (-2.5,1);
\draw[->] (-2.5,0.5) node[right]{$R$} -- (-5,0.5);
\draw[->] (2.5,0.5) node[left]{$T$} -- (3.5,0.5);
\end{tikzpicture}
\caption{The setup for scattering in one dimension. In blue, the situation without scatterer, used in the definition of time delay, is sketched.}
\label{fig:1d}
\end{figure}

We consider one-dimensional scattering and
assume a potential of finite range, i.e., non-vanishing only in a finite interval $[-L, L]$. We choose the position space wave functions of the states $\ket{E}$ outside of the support of the potential as
\beq
\label{eq:x_E_1d}
 \braket{x}{E} = \sqrt{\frac{m}{2 \pi k}} \begin{cases} e^{i k x} + R(E) e^{- i k x} & x < -L \\ T(E) e^{i k x} & x > L, \end{cases}
\eeq
again with $k$ given by \eqref{eq:def_k}.
Here $R, T$ are the usual reflection and transmission coefficients. In the case of a vanishing potential, we have $T= 1$, $R = 0$. Note that the incoming part of the wave function does not depend on the scattering potential, providing a natural way to compare occurrence times for different potentials, i.e., time delays.\footnote{Recall that, when forming wave packets, only the incoming part of the wave function contributes in the asymptotic past $t \to - \infty$, due to stationary phase arguments.} Figure~\ref{fig:1d} provides a sketch of the setup.

We now place a detector of thickness $\delta$ at $x = D > L$, i.e., we are considering the projector represented by the multiplication operator
\beq
 P = \chi_{[D, D + \delta]}(x)
\eeq
on position space wave functions. For the matrix elements of $P$ in the $\ri$ basis, we thus obtain
\begin{align}
 P(E', E) & = \frac{m}{2 \pi \sqrt{k k'}} \bar T(E') T(E) \int_{D}^{D+\delta} e^{ i (k - k') x} \ud x \nn \\
 & = \frac{m \delta}{2 \pi \sqrt{k k'}} \bar T(E') T(E) e^{ i (k - k') D} + \order(\delta^2).
\end{align}
In order to abstract as much as possible from the details of the detector, we consider in the following the limit $\delta \to 0$. Hence, we obtain
\beq
 c^P(E', E) = e^{i ( \arg T(E) - \arg T(E'))} e^{i (k - k') D}.
\eeq
For the corresponding time operator, this implies that
\beq
 t^P(E) = \frac{m}{k} D + \del_E \arg T(E),
\eeq
where we used that $\del_E =\del_E k \del_k = \frac{m}{k} \del_k$.
As expected, the expected arrival time at the detector grows linearly with the distance $D$ of the detector, with coefficient the inverse velocity $\frac{m}{k}$.

A natural application for the time delay between two evolutions (described in the previous section), is now to consider the time delay \wrt the free time evolution (absence of a potential). As described above, we match the energy eigenstates $\ket{E}$ for the two evolutions by identifying the ``in'' part, i.e., we ``identify'' the energy eigenstates \eqref{eq:x_E_1d} and \eqref{eq:x_E_free}. In other words, we identify two states for the evolution with and without potential if their decompositions \eqref{eq:psi_E} \wrt these bases have the same wave function $\psi(E)$.
Obviously, we obtain
\beq
 \bra{\Psi} ( T^P - T_0^P ) \ket{\Psi} = \int | \psi(E) |^2 \del_E \arg T(E) \ud E,
\eeq
with $T_0^P$ the time of occurrence of the event for the free time evolution.
For a wave function sharply peaked at energy $E$, we thus obtain the time delay
\beq
 t_\delay = \del_E \arg T(E).
\eeq

In the case that the potential vanishes for $x < 0$ and has a step at $x = 0$, one classically expects reflection of the particle if the energy is lower than the step height. Classically, this reflection occurs instantaneously, but quantum mechanically there may be a time delay between ``absorption'' and ``emission''. This is thus a natural application of the time delay between two events, as discussed at the end of the previous section. To implement this, we have to set up observables that are selective for the sign of the momentum. For this, we follow the proposal of \cite{ BrunettiFredenhagenTimeObservable} and define
\beq
 P_\pm \defeq \left( \frac{-i \del_x}{2 \sqrt{2 m H}} \pm \frac{1}{2} \right)^* \chi_{[- D -\delta, -D]}(x) \left( \frac{-i \del_x}{2\sqrt{2 m H}} \pm \frac{1}{2} \right).
\eeq
The operator in brackets projects on positive/negative wave numbers, i.e., momenta, so that $P_\pm$ is a model for a detector for a particle within the spatial interval $[-D - \delta, -D]$ and positive/negative momentum. Hence, $P_+$ detects the ingoing and $P_-$ the reflected particle. $P_\pm$ is not a projector, but a positive operator, which is sufficient for the definition of the corresponding POVM describing the distribution of occurrence times \cite{BrunettiFredenhagenTimeObservable}. Analogously to the above, one computes
\begin{align}
 P_+(E', E) & = \delta e^{- i (k - k') D} + \order(\delta^2), &
 P_-(E', E) & = \delta \bar R(E') R(E) e^{i (k - k') D} + \order(\delta^2).
\end{align}
This entails
\begin{align}
 t^{P_+}(E) & = - \frac{m}{k} D, &
 t^{P_-}(E) & = \frac{m}{k} D + \del_E \arg R(E).
\end{align}
In particular, for the case of a wave function peaked at energy $E$, we obtain for the difference of the time of detection of the reflected and the incident particle,
\beq
 t_\delay = \frac{m}{k} 2 D + \del_E \arg R(E).
\eeq
We obtain the classical result $\frac{m}{k} 2 D$ (representing the time needed to travel from the detector at $x = -D$ to $x = 0$ and back) plus a quantum correction. 
A concrete example would a step potential of height $V_0$ for which one straightforwardly finds
\beq
  t_\delay - \frac{m}{k} 2 D = \frac{m}{k} \frac{2}{q}
\eeq
with $q = \sqrt{2 m (V_0 - E)}$. The \rhs has the natural interpretation as 2 times the penetration depth $q^{-1}$, divided by the velocity $\frac{k}{m}$.

\section{Scattering at a spherical potential}
\label{sec:3d}

In the case of scattering at a spherical potential (of short range), the scattering energy eigenstates $\ket{E}$ have position space wave functions which for large distances $r \to \infty$ are given by
\beq
\label{eq:WaveFunctionSpherical}
 \braket{ \bf r}{E} = \sqrt{\frac{m k}{(2 \pi)^3}} \left( e^{i k r \cos \theta} + A(E, \theta) \frac{e^{i k r}}{r} \right) + \order(r^{-2}),
\eeq
with $\theta$ the scattering angle and $A(E, \theta)$ the scattering amplitude. 
Note that smearing the wave functions \eqref{eq:WaveFunctionSpherical} in energy is not sufficient to obtain a normalizable wave function, one also should smear over the direction of the incoming momentum. In the following, we ignore this subtlety, which only becomes relevant if one is interested not just in the expectation value of arrival times but in their distribution, in which case the precise form of the wave function needs to be known.

A detector for the scattered particle at scattering angle\footnote{Scattering in the forward direction is included in the treatment of \cite{Nussenzveig72}, which is based on the dwell time within a ball around the center of the scattering potential.} $\Theta \neq 0$ and distance $D$ can now be modelled by the operator\footnote{On the half-line $\R_+$, the operator $- i \del_r$ is not self-adjoint, but, supplemented with appropriate boundary conditions, symmetric. The operator $P$ is not a projector, only a positive operator. But this is sufficient for the definition of the corresponding POVM describing the distribution of occurrence times \cite{BrunettiFredenhagenTimeObservable}.}
\beq
 P = \frac{1}{1 - \cos \theta} \left( \frac{-i \del_r}{\sqrt{2 m H}} - \cos \theta \right)^* \chi_{[D, D+\delta]}(r) \chi_{[\cos \Theta, \cos \Theta + \eps]}(\cos \theta) \frac{1}{1 - \cos \theta} \left( \frac{-i \del_r}{\sqrt{2 m H}} - \cos \theta \right).
\eeq
One straightforwardly checks that, up to corrections of $\cO(r^{-2})$, this operator annihilates the first term in \eqref{eq:WaveFunctionSpherical}.
 In the limit $\delta, \eps \to 0$, one obtains, in complete analogy to the calculation for the one-dimensional case in the previous section,
\beq
\label{eq:AngularTime}
 t^P(E) = \frac{m}{k} D + \del_E \arg A(E, \Theta).
\eeq

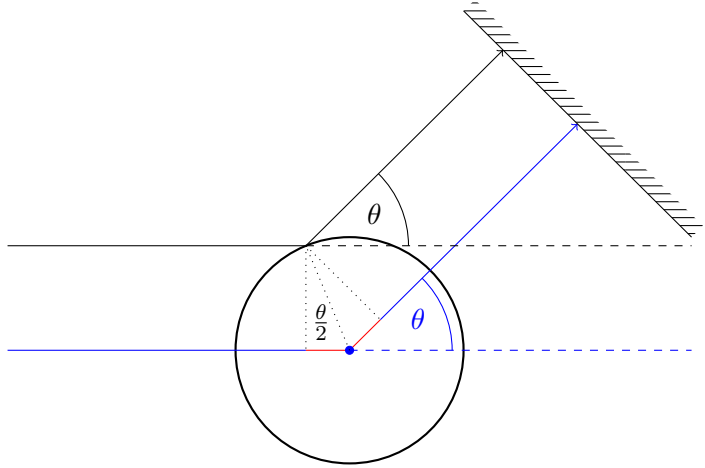
\begin{figure}
\centering
\begin{tikzpicture}[scale=1.5]
 \draw[thick] (0,0) circle (1cm);
 \draw[blue] (-3,0) -- (-0.383,0);
 \draw[red] (-0.383,0) -- (0,0) -- (0.383/1.414,0.383/1.414);
 \draw[blue,->] (0.383/1.414,0.383/1.414) -- (2,2);
 \draw[fill,blue] (0,0) circle (1pt);
 \draw[->] (-3,0.924) -- (-0.383,0.924) -- (2-0.383-0.271,2.924-0.271);
 \draw[dashed] (-0.383,0.924) -- (3,0.924);
 \draw (0.9-0.383,0.924) arc (0:45:0.9);
 \draw (0.6-0.383,0.924+0.1) node[above]{$\theta$};
 \draw[dashed,blue] (0,0) -- (3,0);
 \draw[blue] (0.9,0) arc (0:45:0.9);
 \draw[blue] (0.6,0.1) node[above]{$\theta$};
 \fill[pattern=horizontal lines] (1,3) -- (3,1) -- (3.1,1.1) -- (1.1,3.1) -- (1,3);
 \draw (1,3) -- (3,1);
 \draw[dotted] (0,0) -- (-0.383,0.924);
 \draw[dotted] (-0.383,0.924) -- (-0.383,0);
 \draw[dotted] (-0.383,0.924) -- (0.383/1.414,0.383/1.414);
 \draw (-0.25,0.25) node{$\frac{\theta}{2}$};
\end{tikzpicture}
\caption{The classical time delay for hard sphere scattering. Shown in blue is the scattering at a point scatterer, which is used as a reference. The time it takes to traverse the two red segments is the time delay \eqref{eq:TD_hardSphere_cl}.}
\label{fig:TD_HardSphere}
\end{figure}

In order to obtain a result which does not depend on the precise form of the wave function $\psi(E)$, we should again consider the difference to a reference occurrence time. As discussed above, there are two options: We may compare to a different time evolution, i.e., a different scattering potential, or with a different detector. For the latter option, one may proceed analogously to the one-dimensional case and set up a detector sensitive to the incoming particles. An alternative would be to consider a detector at a reference angle $\theta_0$, leading to the time delay \eqref{eq:t_delay_reference_angle}.
Regarding the first case, i.e., a reference time evolution, we have the difficulty that in the absence of a scattering potential no scattering occurs, so that, in contrast to the one-dimensional case, the free time-evolution does not seem to be a good reference. The problem is already present in the classical case, so one might first wonder how to define a time delay there. It seems that a point scatterer situated at the ``center'' of the scattering potential is the most canonical choice for a reference, as otherwise one would have to specify a reference scattering potential, which seems highly ambiguous.
In the case of scattering off a hard sphere of radius $R$, the classical time delay would then be given by
\beq
\label{eq:TD_hardSphere_cl}
 t_\delay^\class = - \frac{m}{k} 2 R \sin \frac{\theta}{2},
\eeq
\cf Figure~\ref{fig:TD_HardSphere} (the negative sign is due to the path of the particle scattered at the sphere being shorter than the one of the particle scattered at the origin).

In the quantum mechanical case, a model for a point scatterer would be the $R \to 0$ limit of a hard sphere. For the latter, the phase shifts $\delta_\ell$ are given by
\beq
\label{eq:delta_ell}
 \delta_\ell(E) = \arctan \frac{j_\ell(k R)}{n_\ell(k R)},
\eeq
with $j_\ell$, $n_\ell$ the spherical Bessel functions. In terms of the phase shifts, the scattering amplitude is given by\footnote{We recall that the two forms are equivalent for $\theta \neq 0$, by the Legendre polynomial completeness relation. While the first form is usually advantageous in numerical evaluations (where a cut-off in $\ell$ is imposed), the second form is simpler to handle in general treatments, such as the stationary phase arguments used in Section~\ref{sec:WKB}.}
\begin{align}
 A(E, \theta) & = \frac{1}{k} \sum_\ell (2 \ell + 1) \sin \delta_\ell(E) e^{i \delta_\ell(E)} P_\ell(\cos \theta) \nn \\
\label{eq:A}
 & = \frac{1}{2 i k} \sum_\ell (2 \ell + 1) e^{2 i \delta_\ell(E)} P_\ell(\cos \theta),
\end{align}
with $P_\ell$ the Legendre polynomials. With the asymptotic form of the spherical Bessel functions for small arguments,
\begin{align}
j_\ell(x) & \sim \frac{2^{\ell + 1} ( \ell + 1)!}{(2 ( \ell + 1))!} x^\ell, &
n_\ell(x) & \sim - \frac{(2 \ell)!}{2^\ell \ell!} x^{-\ell-1},
\end{align}
we see that, for small $R$,
\beq
 \delta_\ell(E) \sim - \frac{(\ell+1)! \ell!}{(2 (\ell+1))! (2 \ell)!} (2 k R)^{ 2 \ell + 1}.
\eeq
In particular, $A(E, \theta)$ is dominated by the s wave contribution for $R \to 0$, i.e.,
\beq
 A \sim - R.
\eeq
As the leading coefficient in the expansion in $R$ is independent of $E$, we obtain
\beq
 \lim_{R \to 0} \del_E \arg A(E, \theta) = 0.
\eeq
Hence, for the hard sphere in the limit $R \to 0$, the second term in \eqref{eq:AngularTime} vanishes. Defining the angular time delay as the difference of the occurrence times between the actual scattering and the scattering at a point scatterer, modelled by a hard sphere in the limit $R \to 0$, we thus find the relation \eqref{eq:GoldbergerFroissartWatson}. We have thus re-derived \eqref{eq:GoldbergerFroissartWatson} without stationary phase arguments and with a clear ``operational'' definition of the time delay (i.e., \wrt scattering at a point).

\section{The space shift}
\label{sec:SpaceShift}

As noted by \cite{FroissartGoldbergerWatson}, also the displacement of the outgoing particles \wrt a radially outing ray, called the \emph{space shift}, can be defined in the context of quantum mechanical scattering theory, via
\beq
\label{eq:SpaceShiftFGW}
 \bf b = - ( \bf e_z - \cos \theta \ \bf e_r )  \frac{1}{k} \frac{\del \arg A}{\del \cos \theta},
\eeq
when the flux is incoming from the $z$ direction (and $\bf e_r$ is the radial direction at the scattering angle $\theta$, i.e., the outgoing direction). It was derived in \cite{FroissartGoldbergerWatson} in parallel with the derivation of \eqref{eq:GoldbergerFroissartWatson}, using wave packets and stationary phase arguments.
In the following, we present a re-derivation of this result, which is considerably shorter and, we think, simpler than the one given in \eqref{eq:GoldbergerFroissartWatson}. It also provides an alternative ``operational'' definition of the space shift, which may be useful.

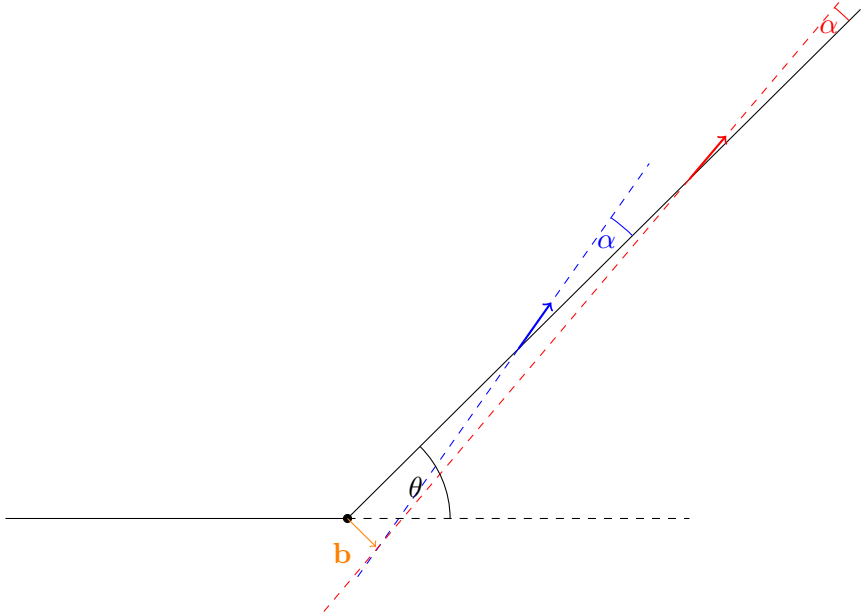
\begin{figure}
\centering
\begin{tikzpicture}[scale=1.5]
 \draw (-3,0) -- (0,0);
 \draw (0,0) -- (4.5,4.5);
 \draw[fill] (0,0) circle (1pt);
 \draw[blue,thick,->] (1.5,1.5) -- ++(55:0.5);
 \draw[blue,dashed] (1.5,1.5) -- ++(55:2);
 \draw[blue,dashed] (1.5,1.5) -- ++(235:2.5);
 \draw[blue] (2.5,2.5) arc (45:55:1.414);
 \draw[blue] (2.1,2.3) node[above right]{$\alpha$};
 \draw[red,thick,->] (3,3) -- ++(50:0.5);
 \draw[red,dashed] (3,3) -- ++(50:2.1);
 \draw[red,dashed] (3,3) -- ++(230:5);
 \draw[red] (4.4,4.4) arc (45:50:1.4*1.414);
 \draw[red] (4.05,4.2) node[above right]{$\alpha$};
 \draw[dashed] (0,0) -- (3,0);
 \draw (0.9,0) arc (0:45:0.9);
 \draw (0.6,0.1) node[above]{$\theta$};
 \draw[orange,->] (0,0) -- (0.25,-0.25) node[midway,below left]{$\bf b$};
\end{tikzpicture}
\caption{Illustration of the space shift. The thick arrows (blue and red) indicate the direction of the probability current at the same angle $\theta$ \wrt the origin (of the scattering potential) and the incoming flux, but at different distances $r$ from the origin. The dashed colored line are the prolongation of these directions. In the limit of large distance $r$, the distance vector of these lines to the origin converges to the space shift $\bf b$ indicated in orange.}
\label{fig:SpaceShift}
\end{figure}

We recall that for a non-normalized wave function $\psi(\bf r)$ (such as a plane wave), the probability current $\bf j(\bf r) = \frac{1}{m} \Imag \bar \psi \nabla \psi$ can be interpreted as the local flux density. For the asymptotic form $\psi_{\mathrm{scatter}}(\vec r) \sim A(\theta) \frac{e^{i k r}}{r}$ of the scattered wave, we obtain the asymptotic form of the current density 
\beq
 \bf j_{\mathrm{scatter}}(\bf r) \sim | A(\theta) |^2 \frac{k}{m r^2} \bf e_r + \Imag (\bar A \del_\theta A) \frac{1}{m r^3} \bf e_\theta.
\eeq
Using that
\beq
\label{eq:del_arg_A}
 \del_\theta \arg A = \Imag \frac{\del_\theta A}{A},
\eeq
this means that the flux of the scattered wave at angle $\theta$ and at large but finite distance $r$ is not exactly radial, but shifted by an angle
\beq
 \alpha = \frac{1}{k r} \del_\theta \arg A, 
\eeq
see Figure~\ref{fig:SpaceShift}. Hence, at given distance $r$ and angle $\theta$ \wrt to the incoming wave (particle current), the origin of the scattered wave seems to be slightly shifted away from the origin. Even though the ``offset angle'' $\alpha$ vanishes as $r \to \infty$, the distance of the corresponding rays (the dashed lines in Figure~\ref{fig:SpaceShift}) to the origin (the center of the scattering potential) converges to the finite value $b = | \frac{1}{k} \del_\theta \arg A |$. Including the direction of the displacement vector, we obtain, for the space shift at scattering angle $\theta$,
\beq
\label{eq:SpaceShift}
 \bf b = - \frac{1}{k} \del_\theta \arg A \ \bf e_\theta,
\eeq
which can easily be seen to coincide with \eqref{eq:SpaceShiftFGW}. For later reference, we define the signed magnitude of the space shift as
\beq
\label{eq:b}
 b \defeq - \frac{1}{k} \del_\theta \arg A.
\eeq
In the situation depicted in Fig.~\ref{fig:SpaceShift}, it is negative, as $\bf b$ points in the direction opposite to $\bf e_\theta$. On the other hand, for the scattering at a hard sphere one straightforwardly reads off 
\beq
\label{eq:b_class}
 b^{\mathrm{class}} = R \cos \frac{\theta}{2}
\eeq
from Figure~\ref{fig:TD_HardSphere} for the space shift in the classical case.

\section{The WKB approximation}
\label{sec:WKB}

For one-dimensional scattering, it is easy to see that the time delay derived from the WKB approximation of the scattering amplitudes coincides with the classical result. This holds both for the case of $E > \max V$ (transmission) and $E < \max V$ (reflection). As we will see, an analogous result holds for the case of scattering at a spherical potential. A similar analysis, but restricted to the case of scattering at small angles, is given in \cite{FroissartGoldbergerWatson}.

The WKB approximation for the scattering phase $\delta_\ell$ of the partial wave of angular momentum $\ell$ is given by \cite{NewtonScatteringTheory}
\beq
 \delta^\WKB_\ell = \frac{\pi}{2} J - \int_{r_0}^\infty \left( \sqrt{2 m (E - V) - J^2 r^{-2}} - k \right) \ud r - k r_0,
\eeq
with $J = \ell + \frac{1}{2}$, the so-called Langer replacement \cite{Langer37}, and $r_0$ the classical turning point at which the square root in the integrand vanishes. Differentiating \wrt the energy yields
\beq
\label{eq:del_E_delta_WKB}
 \del_E \delta^\WKB_\ell = - \int_{r_0}^\infty \left( \frac{m}{\sqrt{2 m (E - V) - J^2 r^{-2}}} - \frac{m}{k} \right) \ud r - \frac{m}{k} r_0
\eeq
The integrand is the difference of the radial velocities in the presence and absence of the potential and the last term gives the time a classical particle needs to reach the origin from distance $r_0$. Hence, this is expression is $\frac{1}{2}$ times the classical time delay at angular momentum $J$.

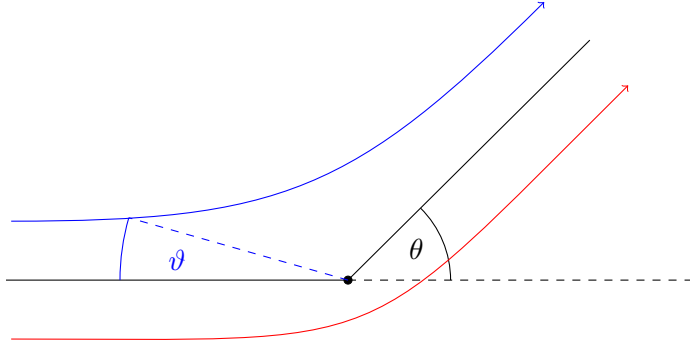
\begin{figure}
\centering
\begin{tikzpicture}[scale=1.5]
 \draw (-3,0) -- (0,0);
 \draw (0,0) -- (45:3);
 \draw[fill] (0,0) circle (1pt);
 \draw[dashed] (0,0) -- (3,0);
 \draw (0.9,0) arc (0:45:0.9);
 \draw (0.6,0.1) node[above]{$\theta$};
 \draw[blue,->] (170:3) to[out=0,in=225,looseness=1.2] (55:3);
 \draw[blue] (-2,0) arc (180:164:2);
 \draw[blue,dashed] (0,0) -- (164:2);
 \draw[blue] (-1.5,0) node[above]{$\vartheta$};
 \draw[red,->] (190:3) to[out=0,in=225,looseness=1.5] (35:3);
\end{tikzpicture}
\caption{Illustration of the scattering angle $\Theta$. For the blue trajectory, $0 < \Theta < \pi$, while for the red trajectory $- \pi < \Theta < 0$.}
\label{fig:Theta}
\end{figure}

Using \eqref{eq:del_arg_A} and \eqref{eq:A}, we have
\beq
\label{eq:del_E_arg_A_sums}
 \del_E \arg A = 2 \Real \frac{\sum_\ell (2 \ell + 1) \del_E \delta_\ell e^{2 i \delta_\ell} P_\ell(\cos \theta)}{\sum_\ell (2 \ell + 1) e^{2 i \delta_\ell} P_\ell(\cos \theta)}
\eeq
In the case when the phase shift $\delta_\ell$ varies slowly with $\ell$, a stationary phase approximation of these sums is meaningful. 
We recall that \cite{NewtonScatteringTheory}
\beq
\label{eq:Theta}
 \Theta(J) = 2 \del_\ell \delta^\WKB_\ell,
\eeq
with $\Theta(J)$ the classical scattering angle at angular momentum $J$. It is here defined as $\Theta \defeq \lim_{t \to \infty} \left( \pi - \vartheta(t) \right)$, where $\vartheta(t)$ is continuous (without $2 \pi$ jumps) and denotes the angle \wrt the incidence axis at time $t$ (so that $\lim_{t \to - \infty} \vartheta(t) = 0$), see Figure~\ref{fig:Theta}. In particular, for a repulsive potential, we have $0 < \Theta < \pi$, while for an attractive potential, we have $- \pi < \Theta < 0$ for trajectories which do not self-intersect, see Figure~\ref{fig:Theta}. For trajectories that ``wind around'' the origin (and thus self-intersect), $\Theta$ can take values smaller than $- \pi$. Hence, the relation to the proper scattering angle $\theta$ is $\theta = | \Theta | \mod \pi$. Furthermore, we recall the large $\ell$ asymptotic
\beq
\label{eq:P_l_asymptotics}
 P_\ell(\cos \theta) \sim \frac{2}{\sqrt{2 \pi \ell \sin \theta}} \cos \left( J \theta - \frac{\pi}{4} \right) 
\eeq
of the Legendre polynomials. Using \eqref{eq:Theta} and \eqref{eq:P_l_asymptotics}, replacing the sums in \eqref{eq:del_E_arg_A_sums} by integrals and using a stationary phase approximation (as in \cite{NewtonScatteringTheory}, Section~18.2.2, for example), one finds that the sums in \eqref{eq:del_E_arg_A_sums} are dominated by the angular momenta $J$ for which $\theta = \pm \Theta(J) \mod 2 \pi$ ($\theta$ is determined modulo $2 \pi$, as in the sums in \eqref{eq:del_E_arg_A_sums} one may insert a factor $e^{2 \pi i n \ell}$ without changing the result). In case there is only one such $J$, one obtains
\beq
 \del_E \arg A \simeq 2 \del_E \delta^\WKB_\ell,
\eeq
where $\ell = J - \frac{1}{2}$, which also establishes the relation of the angular time delay and the Eisenbud-Wigner time delay \eqref{eq:Eisenbud}. By the argument given below \eqref{eq:del_E_delta_WKB}, this corresponds to the classical time delay. If several angular momenta $J$ contribute to scattering at angle $\theta$, then interferences may be relevant, so that the result is in general not simply the weighted sum of the classical time delays of the contributing angular momenta (see also the related discussion for the differential cross section in Section~18.2.2 of \cite{NewtonScatteringTheory}).

To determine the space shift in the WKB approximation, we may use the relation
\beq
 \del_\theta P_\ell(\cos \theta) = \left( \ell + 1 \right) \left( \frac{1}{\sin \theta} P_{\ell + 1}(\cos \theta) - \frac{\cos \theta}{\sin^2 \theta} P_\ell(\cos \theta) \right).
\eeq
Assuming again that only a single angular momentum $J$ contributes (classically) to scattering at scattering angle $\theta$, one finds, using \eqref{eq:P_l_asymptotics}, in a stationary phase approximation,
\beq
 \frac{\del_\theta A}{A} \sim \sqrt{\ell ( \ell + 1)} \frac{e^{\mp i \theta}}{\sin \theta} - (\ell + 1) \frac{\cos \theta}{\sin^2 \theta},
\eeq
where the negative sign in the exponent applies to the case where $\theta = \Theta \mod 2 \pi$, while the positive sign applies to the case $\theta = - \Theta \mod 2 \pi$. 
Hence, we obtain for the signed magnitude \eqref{eq:b} of the space shift, taking into account $\sqrt{\ell (\ell + 1)} = J + \cO(\ell^{-1})$,
\beq
 b \simeq \pm \frac{J}{k} \quad \text{ for } \quad \theta = \pm \Theta \mod 2 \pi.
\eeq
Its modulus is thus the classical impact parameter, as expected. Also the sign is as expected. For example, the red curve in Figure~\ref{fig:Theta}, for which $\theta = - \Theta$, corresponds to the case sketched in Figure~\ref{fig:SpaceShift}, i.e., following the trajectory of the outgoing particle backwards, one passes around the origin. As already discussed for the case sketched in Figure~\ref{fig:SpaceShift}, we have negative $b$, consistent with the above.

\section{Application to the hard sphere}
\label{sec:Examples}

\begin{figure}
\centering
\includegraphics{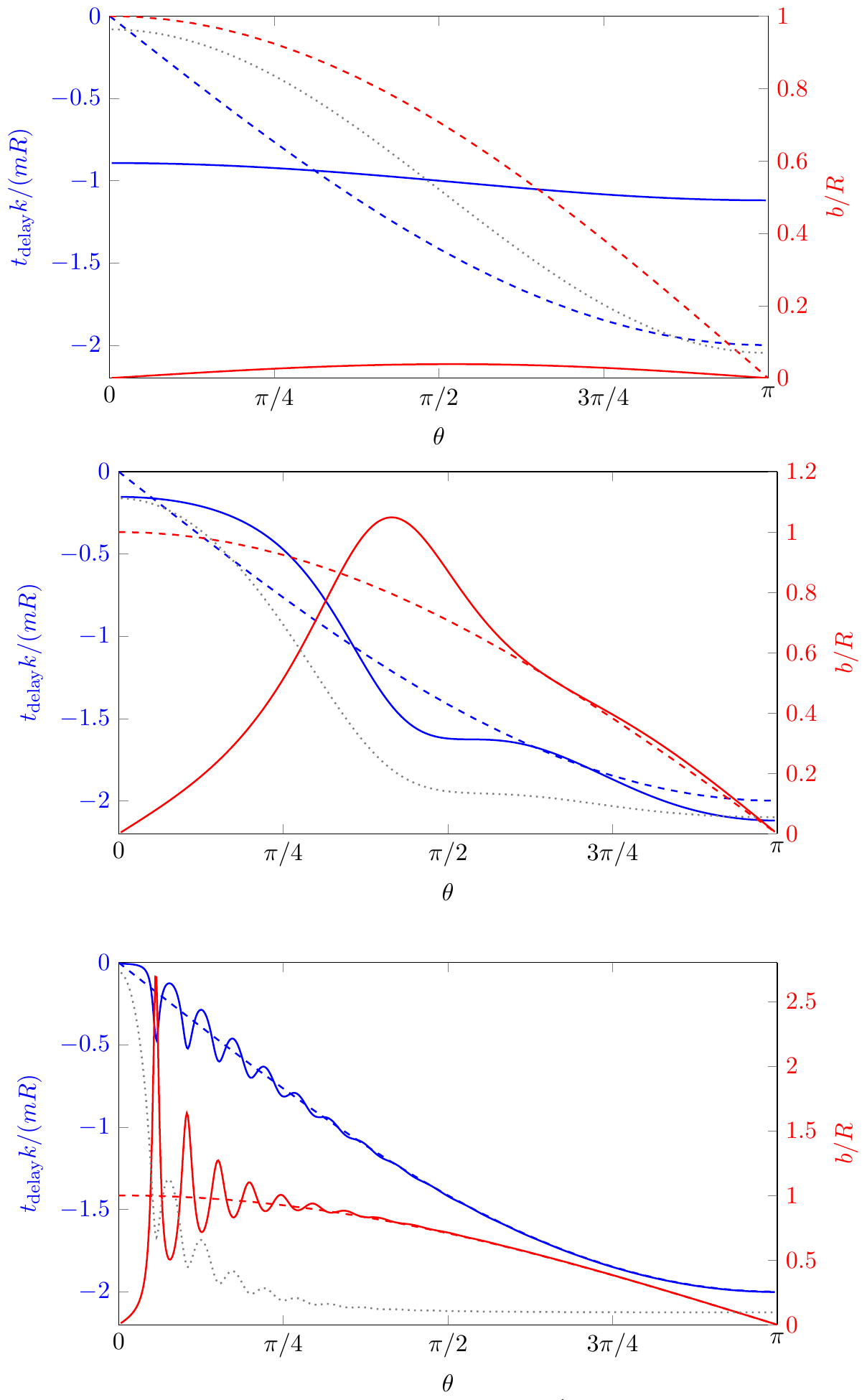}
\caption{Time delay (blue) and the space shift (red) as a function of the scattering angle for scattering at a hard sphere for $k R = 0.2$, $k R =2$, and $k R = 20$. Note that the time delays were multiplied with $k$ so that the same scale for the time delay can be used in all cases. The blue and red dashed lines show the classical time delay \eqref{eq:TD_hardSphere_cl} and space shift \eqref{eq:b_class}. Also indicated (dotted gray) is the differential cross section on a logarithmic scale (not shown).}
\label{fig:HardSpherePlots}
\end{figure}

In Section~\ref{sec:3d}, we already introduced the hard sphere. In particular, we derived the phase shifts $\delta_\ell$, \cf \eqref{eq:delta_ell}, from which the scattering amplitude \eqref{eq:A} can be determined. It is then rather straightforward to evaluate time delay and space shift.\footnote{A plot for the angular time delay in the (not too) short wavelength regime was already shown in \cite{ChaoSkodje}.} Results for different values of $kR$, i.e., different wavelengths, are shown in Figure~\ref{fig:HardSpherePlots}. We see that for large wavelength, $k R < 1$, the time delay is essentially angle-independent and the space shift very small, as expected for $s$ wave dominated scattering. In particular, both deviate substantially from the classical results, as expected in this deep quantum regime. Going to shorter wavelengths, both the time-delay and the space shift approach the classical behaviour, with oscillations (at small angles) that are in phase with the oscillations of the differential cross section (shown as the dotted gray curve on a logarithmic scale). This can be understood intuitively: Assuming an essentially constant velocity of the scattering amplitude $A(\theta, E)$ in the complex plane (as a function of the energy and the scattering angle), one expects that the ``angular velocity'' in the complex plane (either \wrt variation of $E$ or $\theta$), i.e., the time delay or the space shift, to become locally extremal when the scattering amplitude passes close to the origin. For the time delay, this behaviour was also discussed in \cite{ChaoSkodje}. We also note that maxima in the space shift correspond to minima in the time delay. This is naively to be expected, as a positive space shift classically corresponds to a ``short cut'' from the source to the detector, such as in the blue curve in Figure~\ref{fig:Theta}.

The fact that in the short wavelength regime ($k R \gg 1$) and at small angles $\theta$ (i.e., in the forward diffraction region) the time delay and the space shift oscillate around the classical result (with the maximum of the space shift being several times greater than the radius of the scattering sphere) raises the question whether this effect could be in principle observable. A practical difficulty is certainly that the most pronounced extrema occur at the local minima of the differential cross section, so the flux at the corresponding angles will be comparatively small. But there are also limitations arising from the uncertainty principle. Let us discuss this for the example of the space shift. As explained above, the space shift could be determined by measuring the angular component (proportional to $\bf e_\theta$) of the momentum of an outgoing particle at scattering angle $\theta$ and distance $r$, related to the angle $\alpha$ and the space shift $b$ by $p_\theta = k \alpha = - k b / r$. In order to measure such a momentum, the resolution $\Delta p_\theta$ of the detector should be smaller, i.e., $\Delta p_\theta \lesssim k | b | / r$. By the uncertainty principle, this corresponds to an angular width of the detector of $\Delta \theta \gtrsim 1/(2 r \Delta p_\theta) \gtrsim 1/(2 k | b |)$. For the peaks in the space shift to be detectable, this angular width must not be greater than the width of the peaks. For the case $k R = 20$ and the most pronounced peak of the space shift, we have $b \simeq 2.7 R$ and a width $\Delta^\theta_{\mathrm{peak}} \simeq 0.03$ (corresponding to fall-off to half the peak height).\footnote{In case one performs relative measurements (for example by comparing the space shifts at different angles, or for different energies), it should rather be the height and width of the peak \wrt the adjacent minima which are relevant. However, as the minima adjacent to the first peak are (close to) zero, this would not substantially alter the conclusions of our discussion.} For the minimal angular width of the detector, we thus obtain $\Delta \theta \gtrsim 0.01$, which is of the same order of magnitude as the width of the peak, but can still be accommodated within. The relation becomes even better for higher energy, as the peak width approximately scales as $k^{-1}$, but the peak height also grows. For example, at $k R = 50$, we have $b \simeq 4.5 R$, corresponding to $\Delta \theta \gtrsim 0.002$, and a peak width $\Delta^\theta_{\mathrm{peak}} \simeq 0.01$. Hence, we conclude that the peaks of the space shift in the forward diffraction pattern should in principle be observable.

A similar reasoning can be applied to the peaks (in negative direction) of the angular time delay. It is now the time-energy uncertainty relation $\Delta t \Delta E \gtrsim \frac{1}{2}$ that limits the observability of the effect, as in order to detect a time delay $t_\delay$, we should use pulses of duration $\Delta t \lesssim | t_\delay |$.\footnote{In this discussion, we neglect the spreading of the wave packet due to dispersion. In order to circumvent this further difficulty, it seems to be advantageous to consider analogous situations in optics, in case one intends to actually verify the effect experimentally.} This means that there must be a spread $\Delta E \gtrsim 1/(2 | t_\delay |)$ in energy. If the width $\Delta^E_{\mathrm{peak}}$ in energy of the peak in $t_\delay$ is now smaller than this spread, the effect is ``washed out'' and thus unobservable. The first ``downward'' peak of $t_\delay$ for $k R = 20$, shown in the third plot of Fig.~\ref{fig:HardSpherePlots}, has height $|  t_\delay | \simeq 0.02 m R^2$, corresponding to $\Delta E \gtrsim 25/(m R^2)$. For the width of the peak (at fixed $\theta$, as a function of the energy), one finds $\Delta^E_{\mathrm{peak}} \simeq 80/(m R^2)$. Again, it seems that the peak in the angular time delay should in principle be observable.

\section{Conclusion}

Using the concept of the time of occurrence of an event \cite{BrunettiFredenhagenTimeObservable}, we re-derived the angular time delay \eqref{eq:GoldbergerFroissartWatson} of \cite{FroissartGoldbergerWatson}. In particular, we clarified the choice of a reference dynamics (scattering at a point) \wrt which the angular time delay is defined. We also re-derived by elementary means (using the probability current instead of a wave packet) the space shift \eqref{eq:SpaceShiftFGW} of \cite{FroissartGoldbergerWatson}. We showed that in the WKB approximation, the expressions reduce to their classical counterparts, at least if only a single trajectory contributes to scattering to a given angle $\theta$ in the classical case. We applied the concepts to the example of the hard sphere and found (as expected) essentially angle independent time delay and space shift in the deep quantum (s wave dominated) regime. We also saw convergence to the classical results in the short wavelength regime at large scattering angle, but observed pronounced oscillations (with the angle) of time delay and space shift in the forward diffraction region. We argued that these oscillations should be in principle observable.

As for future developments, one might, on the conceptual side, consider generalizations to relativistic scattering theory. Regarding applications, we saw in the example of scattering at the hard sphere that there are strong variations (with the scattering angle) of time delay and space shift in the forward diffraction region (in particular pronounced peaks at the local minima of the differential cross section, i.e., the ``dark'' areas in the diffraction pattern). We argued that these should be in principle observable. It is plausible that a similar effect occurs in optical diffraction (at a single slit, for example), which may pave the way for experimental conformation.

%\bibliographystyle{../h-elsevier_new}
%\bibliography{../mybib}

\end{document}